\documentclass[aps,preprint,showpacs,amsmath,amssymb,superscriptaddress]{revtex4-1}
\newcommand{\fant}[1]{\phantom{#1}}
\newcommand{\be}{\begin{equation}}
\newcommand{\ee}{\end{equation}}
\newcommand{\wdg}{\wedge}
\newcommand{\ot}{\otimes}
\usepackage{multirow}

\begin{document}
\title{Generalized Sparling-Thirring  form in the Brans-Dicke theory}

\begin{abstract}
The definition of the Sparling-Thirring form is extended to the Brans-Dicke theory. By writing the Brans-Dicke field equations in a formally Maxwell-like form, a superpotential and a corresponding pseudo ener\-gy-momentum form are  defined. The general energy expression provided   by the superpotential in the Jordan frame is discussed in relation to the corresponding expression in the Einstein frame. In order to substantiate its formal definition,  the generalized Sparling-Thirring  form is used to calculate the energy for the spherically symmetric vacuum solution in the Brans-Dicke theory.

\end{abstract}


\author{Ahmet Baykal}
\email{abaykal@nigde.edu.tr}
\affiliation{Department of Physics, Faculty of Arts and Sciences, Ni\u gde University,  Bor Yolu,  51240 Ni\u gde, Turkey}
\author{\"Ozg\"ur Delice}
\email{ozgur.delice@marmara.edu.tr }
\affiliation{Department of Physics, Faculty of Arts and Sciences, \\Marmara University, 34722 \.Istanbul, Turkey}
\date{\today}

\maketitle

\section{Introduction}

According to the equivalence principle, the gravitational field, as described by Einstein as the geometry of spacetime, can locally be transformed away  at a given point so that a curved spacetime looks locally like  a flat Min\-kows\-kian at an infinitesimal level.
Conversely, an accelerated frame in flat spacetime can emulate gravity \cite{thirring}.  Consequently, one can argue that the equivalence principle  renders any definition of a local energy density to be non-tensorial.  This line of thought lead to the definition of pseudo energy-momentum forms that depend on a chosen frame. In particular, for an asymptotically flat  spacetime, a working definition of energy-momentum can be provided.
A well-known example of such a definition of pseudo-energy momentum  form and the superenergy,  in the framework of the general relativity theory (GR), is the Sparling-Thirring 2-form and the associated pseudo-energy momentum forms \cite{thirring,wallner-thirring,sparling}.
In GR, the Sparling-Thirring form is introduced to define the total energy and it is related to the definition of energy by Arnowitt-Deser-Misner
(known as the ADM energy in short). The Sparling-Thirring form can be related to  other definitions of energy, such as the well-known Landau-Lifschitz energy \cite{wallner-thirring,szabados,benn,waldyr}.

The constructions  of a pseudotensor using field equations of gravity and the definitions involving
the Noether symmetries have previously been discussed in connection with the quasilocal quantities in \cite{nester,nester2,obukhov-rubilar1,obukhov-rubilar2,obukhov-rubilar3}.
In a completely different approach, but using a similar mathematical notation and set up, the Bell tensor is derived from a variational principle in the context of a particular tensor-tensor model of gravitation  \cite{dereli-tucker-tensor-tensor}. More recently, conserved Noether quantities were also studied  in the context of  the motion of spinning particles in the BD theory \cite{burton-tucker-wang}. A general formulation of generalized Bianchi identities relative to
conformally related frames and corresponding conservation laws in various modified gravity models have been discussed  in \cite{koivisto}.

The scalar-tensor theory of gravitation  proposed by Brans and Dicke \cite{BD-original} almost half a century ago is still one of the popular
theories of gravity which received rekindled interest in the applications in particular to the cosmological dynamics as
possible alternatives to dark energy and dark matter \cite{Faraonibook,fujui-maeda}.  The dynamical equivalence   between  $f(R)$-type modified gravity models to the scalar tensor theories,  introduced  in \cite{whitt,teyssandier} for particular modified gravity models, also increased  the interest to  these theories \cite{clifton-phys-rep,Nojiri,faraonireview}. For a treatment of this equivalence in the differential form calculus see {\cite{baykal-delice2}. Any modified theory of gravity should compete with
the GR on the observational front at cosmological and solar system scale. On the theoretical side, a prediction of such a modified theory  is to be compared to those of GR. The conserved currents for BD theory has been studied by applying the  Lagrange-Noether machinery \cite{hart1,hart2,barraco-hamity}. More recently, using the exterior algebra of differential forms, and in a more general context of Noether's theorem, a new conserved charge is defined as an  application in \cite{obukhov-PLB}. In an earlier work, the Landau-Lifshitz superpotential is extended to BD theory \cite{nutku}.

In the approach making use of the field equations,
the definition of a pseudo-tensor can be applied to  any geometrical theory  that follow from any Lagrangian once the field equations formulated in terms of the tensor-valued differential forms. For example,  a generalized Sparling-Thirring form was previously  defined  \cite{madore} for
the dimensionally continued Euler-Poincare form
\be\label{lagrangian-EP}
L_{EP}
=
\Omega_{ab}\wdg\cdots\wdg\Omega_{cd}*e^{ab}\wdg\cdots \wdg e^{cd},
\ee
where $L_{EP}$ can be considered as natural generalization of Einstein-Hilbert  Lagrangian
\be
L_{EH}
=
\frac{1}{2\kappa^2}\Omega_{ab}\wdg*e^{ab},
\ee
leading to second order field equations (in the notation is introduced in the following section).
In this generalization, an exact differential is separated out from the field equations following from  (\ref{lagrangian-EP}) using the definition
of the curvature 2-forms as in the case of the Einstein field equations.

In the present work, the Sparling-Thirring construction is extended to the BD scalar-tensor theory.  The  BD scalar field is incorporated into the definition of Sparling-Thirring form, as a  consequence of the  vanishing torsion constraint on the connection 1-form obtained by  the constrained first order formalism to the BD theory Lagrangian.

The organization of the paper is as follows. In the next section, after briefly discussing the notation of the tensor-valued forms and  the algebra of differential forms, the  BD field equations are obtained by the variation of the BD action via a first order formalism. The use of the modern language of differential forms has several advantages over  tensorial treatment, in that the differential forms are natural integrands on manifolds and allows one to use Stokes' theorem to convert volume integrals to flux integrals. Moreover, the use of differential forms also allow one to define energy for asymptotically flat spacetimes without introducing an approximation scheme. The definition of energy superpotential is therefore requires the BD field equations to be formulated relative to an orthonormal coframe. Due to this, we express the well known BD field equations relative to an orthonormal coframe in this section.
In section 3,  following the construction of Sparling-Thirring form in GR, the BD field equations in this form is used to define a superpotential and a corresponding energy momentum form which involves the nonminimally coupled dynamical BD scalar field as well. In accordance with the equivalence principle, the resulting expressions for energy and momentum  depends on a chosen frame as in GR.
Since the definition of BD pseudo-energy-momentum form  is particularly suitable  to discuss the mathematical expressions in Jordan frame and conformally related Einstein frames, we obtain relevant expressions in the Einstein frame as well in section 4. In section 5, the energy definition is used to define the total mass of the spherically symmetric and asympto\-tically flat vacuum BD solution. The paper ends with general remarks regarding to some possible  energy de\-fi\-nitions in related modified gravity theories
in line with the discussion below.

\section{BD Field equations}
The notation used for the algebra of differential forms is adopted from \cite{benn-tucker}.
The calculations in this paper will be carried out relative to a set of  orthonormal basis coframe 1-forms $\{e^a\}$ for which the metric reads
$g=\eta_{ab}\, e^a\ot e^b$ with $\eta_{ab}=\mbox{diag}(-+++)$. The set of basis frame fields is $\{X_a\}$ and the abbreviation $i_{X_a}\equiv i_a$ is used for the contraction operator with respect to the basis frame field $X_a$. $*$ denotes the Hodge dual operator acting on the basis forms and
$*1=e^{0}\wdg e^1\wdg e^2\wdg e^3$ is the oriented volume element. The abbreviations of the form $e^{a}\wdg \cdots \wdg e_c\wdg \cdots\equiv
e^{a\cdots}_{\fant{aaa}c\cdots}$ for the exterior products of basis 1-forms are  extensively used for the convenience of the notation.
The first structure equations  of Maurer-Cartan 
 read
\be\label{cse1}
\Theta^a
=
De^a
=
de^a+\omega^{a}_{\fant{a}b}\wdg e^b=0,
\ee
with the  vanishing torsion two form, i. e.,  $\Theta^a
=0$.
$D$ is the covariant exterior derivative operator, acting on tensor-valued forms, and a suitable definition and its relation to covariant derivative can be found for example in \cite{thirring}.
The curvature 2-form $\Omega^{a}_{\fant{a}b}$ with
$\Omega^{a}_{\fant{a}b}=\frac{1}{2}R^{a}_{\fant{a}bcd}e^{cd}$ satisfies the second structure equation of the Maurer-Cartan
\be
\Omega^{a}_{\fant{a}b}
=
d\omega^{a}_{\fant{a}b}
+
\omega^{a}_{\fant{a}c}\wdg \omega^{c}_{\fant{a}b}.\label{secondstr}
\ee

The discussion below will be based exclusively on the original BD Lagrangian among other scalar-tensor theories.
The approach to the definition of a total energy in BD theory below involves surface integrals of pseudo-tensorial
quantities, therefore it is convenient to use the algebra of differential forms defined on pseudo-Riemannian manifolds
that are already built for integration.

Expressed in terms of differential forms,  the total Lagrangian 4-form for the original BD Lagrangian with matter fields included has the form
\be\label{BD-lag}
L_{tot.}
=
L_{BD}[\phi, e^a, \omega^{a}_{\fant{a}b}]+\frac{8\pi}{c^4}L_{matter}[g,\psi],
\ee
where the gravitational part  in the so-called Jordan frame is given by
\be
L_{BD}
=
\frac{\phi}{2} \Omega_{ab}\wdg *e^{ab}-\frac{\omega}{2\phi}d\phi\wdg *d\phi,
\ee
and  the matter part $L_{matter}[g, \psi]$ is assumed to be independent of the BD scalar and the spinor fields.
The gravitational coupling constant $G$ in GR is replaced by a dynamical scalar field $\phi^{-1}$ with a corresponding kinetic term for the scalar field.
$\omega$ is the free BD parameter and GR limit is recovered in the $\omega\mapsto\infty$ limit. For the vacuum solutions or for when the matter energy momentum tensor has a  non vanishing trace, however, things are more complicated and the above limit may not give \cite{romero-barros,banerjee-sen,faraoni-w-dependence} the GR limit, unless the arbitrary parameters are fixed by some physical grounds such as post Newtonian extension for asymptotically flat solutions \cite{bhadra-nandi}.

In the general framework of first order formalism for gravity, the independent gravitational variables can be taken as  the set of basis coframe 1-forms $\{e^a\}$ and the connection 1-forms $\{\omega^{a}_{\fant{a}b}\}$. The local Lorentz invariance forbids any gravitational action to have explicit dependence on $\{\omega^{a}_{\fant{a}b}\}$ and the first order derivatives of $de^a$ and $d\omega^{a}_{\fant{a}b}$ enters into a gravitational Lagrangian
with local Lorentz symmetry by the tensorial quantities such as $\Theta^a$ and $\Omega^{a}_{\fant{a}b}$. The minimal coupling prescription for the matter fields
implies that $de^a$ and $d\omega^{a}_{\fant{a}b}$ occurs only in the gravitational sector in the presence of the matter fields. On the other hand,
the BD scalar field  $\phi$  couples to the metric nonminimally.
Because of the nonminimal coupling, the BD scalar  field is dynamical even in the absence of the kinetic term for it.

An essential ingredient of the discussion is the formulation of the BD equations in terms of exterior forms which then allows one to
write the conservation laws in differential and integral forms as well easily with the help of Stokes' theorem with considerable technical advantage over
the methods using tensor components. Hence, a derivation of BD equations in the desired form using exterior algebra of forms using the Lagrange multiplier method is  presented below.

The vanishing torsion constraint can be implemented into the variational procedure by introducing Lagrange multiplier 4-form term
\be
L_C
=
\lambda_a\wdg \Theta^a,
\ee
to the original Lagrangian form $L_{BD}$, where the Lagrange multiplier 2-form $\lambda_a$ is a vector-valued 2-form imposing the dynamical constraint $\Theta^a=0$. The Lagrangian $L_{ext.}=L_{ext.}[\phi, e^a, \omega^{a}_{\fant{a}b}, \lambda_a, \psi]$  for the extended gravitational part then has the explicit form
\be
L_{ext.}
=
L_{BD}[\phi, e^a, \omega^{a}_{\fant{a}b}]+L_C[e^a,  \omega^{a}_{\fant{a}b}, \lambda^a].
\ee
The total variational derivative of $L_{ext.}$ with respect to independent variables can be found as
\begin{eqnarray}
\delta L_{ext.}
&&=
\delta\phi
\left(
\frac{1}{2}R*1
-
\frac{\omega}{2\phi}d*d\phi
+
\frac{\omega}{2\phi^2}d\phi\wdg *d\phi
\right) \nonumber
+
\delta e_a\wdg
\left(
\frac{\phi}{2}\Omega_{bc}\wdg *e^{abc}
+
D\lambda^a
+
\frac{\omega}{\phi}*T^a[\phi]
\right)
\nonumber\\
&&
+
\delta\omega_{ab}\wdg
\frac{1}{2}
\left[
D(\phi*e^{ab})
-
(e^a\wdg \lambda^b-e^b\wdg \lambda^a)
\right]
+
\delta\lambda_a\wdg \Theta^a 
\label{total-var-der},
\end{eqnarray}
up to an omitted boundary term.
The energy-momentum 3-form of the scalar field $*T^a[\phi]=T^a_{\fant{a}b}[\phi]*e^b$ has the explicit expression
\be
*T^a[\phi]
\equiv
\frac{1}{2}\left[(i_ad\phi)*d\phi+d\phi\wdg i_a*d\phi\right].
\ee

Assuming that there are no spinor fields to couple  to the connection, the independent connection equations read
\be
D(\phi *e^{ab})
-
(e^a\wdg \lambda^b-e^b\wdg \lambda^a)
=0.
\ee
These equations can be considered as the defining equation for the Lagrange multiplier 2-forms $\lambda^a$ and they can  uniquely be solved for the Lagrange multiplier 2-forms by calculating its contractions by taking  the constraint $\Theta^a=0$ into account. One finds that
\be\label{lag-mult-expression}
\lambda^a
=
*(d\phi\wdg e^a).
\ee

Consequently, using the expression (\ref{lag-mult-expression}) for the Lag\-range multiplier form in the metric field equations induced by the coframe variational derivative  $\delta L_{ext.}/\delta e^a \  \equiv -*E^a=0$ in (\ref{total-var-der})  read
\be\label{BD-eqns1}
*E^a
=
\phi *G^a-D*(d\phi\wdg e^a)-\frac{\omega}{\phi}*T^a[\phi],
\ee
where $E^a\equiv E^a_{\fant{a}b}e^b$ is vector-valued 1-form.
In the pre\-sence of the matter fields,  the BD field equations take the form
\be
*E^a-\frac{8\pi}{c^4}*T^a[\psi]=0,
\ee
where $*T^a[\psi]$ stands for the energy-momentum form for the matter field $\psi$ derived from the variational derivative of $L_{matter}[g,\psi]$ with respect to the basis coframe 1-forms. $*T^a[\psi]$ depends on the metric tensor as well and therefore in the notation used here, it may also be appropriate to state the dependence by writing it in the form $*T^a[\psi,e^a]$. However, the metric dependence is surpassed for simplicity in the case that the metric dependence is obvious from the context in the discussions relative to a given coframe.

Although it is customary to write the BD equations (\ref{BD-eqns1}) by dividing it with $\phi$, the Lagrange multiplier and the
${\omega}{\phi^{-1}}*T^a[\phi]$ terms are considered to be on the ``geometry" side of the BD field equations.
As a consequence of the diffeomorphism invariance of the BD Lagrangian, it follows from the corresponding Noether  identity that $D*E^a=0$ \cite{hehl-mccrea-mielke-neemann}.
Alternatively, in a more straightforward way, and with the help of the identities

\begin{eqnarray}
&&D\left(\phi*G^a\right)
=
d\phi\wdg *G^a,\nonumber
\\
&&D^2*(d\phi\wdg e^a)
=
\Omega^{a}_{\fant{a}b}\wdg *(d\phi\wdg e^b)
=
d\phi \wdg*R^a,
\\
&&D\left(\frac{\omega}{\phi}*T^a[\phi]\right)
=
\frac{\omega}{2\phi}(i^a d\phi)\left(d*d\phi-\frac{1}{\phi}d\phi\wdg *d\phi\right),\quad\nonumber
\end{eqnarray}
it is possible to arrive at  the identity
\be
D*E^a
=
-\frac{1}{2}(i^ad\phi)\bigg(  \frac{\omega}{\phi}\,d*d\phi  -\frac{\omega}{\phi^2}d\phi\wdg *d\phi+R*1\bigg),\
\ee
as expected. The right-hand side vanishes identically provided that the field equation for the BD scalar is satisfied since the terms on the right-hand side
inside the bracket are proportional to the field equations for the BD scalar given below.
This consideration is in line with the well-known case of the Einstein-massless scalar field equations,  $R_a=(i_{a}d\phi)d\phi$, from which the field equations for $\phi$ follows from the Bianchi identity.
Consequently, one has $D*T^a[\psi]=0$, from which one can derive geodesic postulate for test particles  by introducing the matter energy momentum
of ideal fluid \cite{benn-tucker}. As it is well-known, this is not the case in the conformally related Einstein frame, where scalar field  couples nonminimally to a matter field.

On the other hand, the field equation for the BD scalar that follows from the variational derivative \\ $\delta L_{BD}/\delta \phi=0$ is given by
\be\label{scalar eqn1}
\omega\, d*d\phi-\frac{\omega}{\phi} \, d\phi\wdg *d\phi+\phi\, R*1=0.
\ee
The BD scalar couples to the matter energy momentum through the last term in (\ref{scalar eqn1}).  Then, by
combining the scalar field equation with the trace of the metric equations,  the equation for the BD scalar reduces  to
\be\label{reduced-scalar-eqn}
d*d\phi
=
\frac{8\pi}{c^4}\frac{1}{2\omega+3}T[\psi]*1,
\ee
for which the trace $T[\psi]\equiv T^{a}_{\fant{a}a}[\psi]$ of the matter energy-momentum  tensor act as the source term.
As pointed out in the invariance of $L_{BD}$ discussion above, the reduced field equation (\ref{reduced-scalar-eqn}) follows  from Bianchi identity for the BD field equations together with its trace.

\section{Definitions of the superpotential and the pseudo-energy-momentum forms}

The definitions of the superpotential for the BD theory can be given in terms of the original Sparling-Thirring 2-form in GR. Thus it is appropriate to
recall the construction of Sparling-Thirring form\cite{thirring,wallner-thirring,sparling}.
The Einstein 3-form is defined in terms of the following contraction
\be\label{eintein-3form-def}
*G^a
=
-\frac{1}{2}\Omega_{bc}\wdg *e^{abc},
\ee
where 1-form $G^a$ can be defined  in terms of the Einstein tensor components as $G^a\equiv G^{a}_{\fant{a}b}e^b$. The definition (\ref{eintein-3form-def})
is suitable to separate out an exact differential out of Einstein tensor 
regardless of the
presence of any matter Lagrangian.
Explicitly, by inserting the second structure equation (\ref{secondstr}) into the right-hand side of (\ref{eintein-3form-def}),  probably technically  in the most straightforward way, one finds
\be
*G^a
=
d*F^a+*t^a,
\ee
with the Sparling-Thirring 2-form
\be\label{thirring-form-def}
*F^a
=
-\frac{1}{2}\omega_{bc}\wdg *e^{abc},
\ee
and the gravitational energy-momentum (pseudo-tensor) 3-form
\be\label{gr-ll-energy}
*t^a
\equiv
\frac{1}{2}
(\omega_{bc}\wedge\omega^{a}_{\phantom{Q}d}
\wedge*e^{bcd}
-
\omega_{bd}\wedge\omega^{d}_{\fant{a}c}\wedge
*e^{abc}).
\ee

The classical definitions of the Sparling-Thirring superpotential and the corresponding energy-momentum pseudo-tensor above can be related to well-known expressions for other  superpotentials \cite{wallner-thirring}. In particular,  the  components $t_{ab}$ of the pseudo-tensorial object $*t^a=t^{a}_{\fant{a}b}*e^b$ are symmetrical relative to a coordinate basis \cite{thirring}.

The definitions (\ref{thirring-form-def}) and (\ref{gr-ll-energy}) then help to split  $*E^a$, in an equally straightforward way as in the GR case above, in the following  Maxwell-like form for BD theory
\be\label{BD-eqns-maxwell-form}
*E^a
=
d*\mathcal{F}^a
+
*\mathcal{T}^a=\frac{8\pi}{c^4}*T^a[\psi],
\ee
with the generalized definitions of the superpotential
\be
*\mathcal{F}^a
\equiv
\phi*F^a-*(d\phi\wdg e^a),
\ee
and the corresponding energy-momentum form
\be\label{pseudo-energy-def}
*\mathcal{T}^a
=
\phi*t^a
-d\phi\wdg *F^a
-
\omega^{a}_{\fant{a}b}\wdg *(d\phi\wdg e^b)
-
\frac{\omega}{\phi}*T^a[\phi],
\ee
for the superpotential and the pseudo-energy-momentum forms respectively for the BD vacuum case.
A  conserved pseudo 4-current then follows from the equations (\ref{BD-eqns-maxwell-form}) as the simple consequence of the differential identity $d^2\equiv0$.
The expression on the right-hand side of (\ref{pseudo-energy-def}) is pseudo tensorial and it involves connection 1-forms which can be transformed away at a given point in accordance with the equivalence principle.

The formulation of the field equations in the form of a conservation law in terms of differential forms allows one to rewrite the conserved quantities as  flux  integrals by making use of Stokes's theorem. By defining total energy-momentum pseudo form
\be
*\tau^a
\equiv
\frac{8\pi}{c^4}*T^a[\psi]-
*\mathcal{T}^a,
\ee
a conserved 4-current $P^a_{BD}$ then can be expressed as the following flux integral
\be\label{4-mom-def}
P^a_{BD}
\equiv
\int_{U}*\tau^a
=
\int_{U}d*\mathcal{F}^a
=
\int_{\partial U}*\mathcal{F}^a,
\ee
with $\partial U$ as the boundary of a three-dimensional submanifold $U\subset M$ on a pseudo-Riemannian manifold $M$.

The BD superpotential $*\mathcal{F}^a$ can be put in a simplified form analogous to the Sparling-Thirring form as
\be\label{BD-superpotential-def}
*\mathcal{F}^a
=
-\frac{1}{2}\Lambda_{bc}\wdg *e^{abc},
\ee
with a modified connection 1-form denoted by $\Lambda_{bc}=-\Lambda_{cb}$ that is given explicitly by the following form
\be\label{Lambda}
\Lambda_{bc}
\equiv
\phi \,\omega_{bc}+\frac{1}{2}(e_b i_cd\phi-e_c i_bd\phi).
\ee
By definition,  1-form $\Lambda_{bc}$ incorporates the constraint on the connection 1-form resulting from nonminimal coupling of the BD scalar
in a peculiar  way.

The BD superpotential $*\mathcal{F}^a$ defined in (\ref{BD-superpotential-def}) naturally involves BD scalar which carries dynamical degrees  of freedom
and the scalar contribution to the superenergy definition results from the vanishing torsion constraint term on the independent connection.
Consequently, the dynamical coupling constant in BD theory also incorporated into the total energy by definition through the Lagrange multiplier term and at the same time the construction is guided by the mathematical structure of the field equations formulated in terms of differential forms. By construction, the conserved quantities require the field equations to be satisfied.

In parallel to the original definition of the total energy in terms of  Sparling-Thirring form
for an asymptotically flat geometry in GR, one can define the total energy as the temporal component of conserved 4-current $E_{BD}\equiv P^0_{BD}$  as
\begin{eqnarray}
E_{BD}
\equiv 
\int_{S^2_\infty}*\mathcal{F}^0 
=
-\frac{1}{2}\int_{S^2_\infty}\left(\phi\omega_{jk}+\frac{1}{2}(e_j i_kd\phi-e_k i_jd\phi)\right)\wdg *e^{0jk},\quad
\end{eqnarray}
where the flux integral is over  two dimensional sphere with infinite radius denoted by $S^2_{\infty}$ evaluated at a cons\-tant value of $t$. As for the other  classical pseudo energy-momentum and superpotential forms, the 4-momentum definition (\ref{4-mom-def}) depends on the frame in which it is computed \cite{nester}. In the same way as the superpotentials are calculated in GR for an asymptotically flat spacetime, the BD superpotential is to be calculated in a coordinate system that is most nearly Minkowskian as well.

In contrast to the previous energy-momentum pseu\-do-tensors for scalar-tensor theories in the literature, the use of exterior forms in
defining conserved currents in BD theory is technically straightforward which amounts to a judicious arrangement of the terms in the field equations.
This arrangement of the field equations not only separates the higher order terms but also singles out the leading terms that are linear in the derivatives of the field variables.

\section{Jordan frame vs. Einstein Frame}

The definition of the total energy in BD theory above involves no approximation scheme exercised in other approaches, for example linearization  of the field equations around a suitable background solution \cite{deser-tekin-construction,cebeci-tekin-sarioglu}. The corresponding definition in the Einstein frame below allow one to discuss the energy definition relative to Einstein frame in a straightforward manner.

It is a well-known fact that the conformal scaling  of the metric components $g_{\mu\nu}$ (relative to a coordinate coframe with Greek indices) by
\be\label{ct-def-metric-comp}
g_{\mu\nu}
\mapsto
\tilde g_{\mu\nu}=\phi\, g_{\mu\nu},
\ee
brings the BD Lagrangian into the Einstein frame with a new scalar field coupled non-minimally to matter fields. The conformal transformation
(\ref{ct-def-metric-comp}) is equivalent to the scaling of the coframe basis 1-forms as \cite{cebeci-dereli-torsion}
\begin{equation}\label{ct-def}
e^a\mapsto
\tilde{e}^a
=
\phi^{1/2} e^a.
\end{equation}
Furthermore, the interior product operators of the frames are related by
$
i_{X_a}= \phi^{1/2} i_{\tilde{X}_a},
$
whereas the invariant volume forms are related by $*1=\phi^{-2}\tilde *1$.
Consequently, the connection 1-forms transform as
\begin{equation}\label{tildeomega}
\omega_{ab}
\mapsto
\tilde \omega_{ab}=\omega_{ab}-\frac{1}{2\phi}\left( e_bi_{a}d\phi  - e_ai_{b}d\phi \right).
\end{equation}	
The conformal transformation (\ref{ct-def}) brings BD Lagrangian into the so-called Einstein frame with a minimally-coupled massless scalar field $\alpha$ as
\begin{equation}
\tilde L_{BD}= \frac{1}{2}\tilde \Omega_{ab}\wdg\tilde *\tilde e^{ab}-\frac{1}{2}d\alpha\wdg \tilde * d\alpha,
\end{equation}
up to an omitted  closed form. The scalar field $\alpha$ is related to $\phi$ by
\begin{equation}
\alpha=(\omega+{3}/{2})^{1/2} \ln\phi.
\end{equation}
The metric field equations that follow from the scaled Lagrangian can be obtained as
\be
\tilde *\tilde G_a-\tilde * \tilde T_a[\alpha]-e^{-2b\alpha}\tilde* \tilde T_a[\tilde g, \psi]=0,\label{FeqnEinsfr}\\
\ee
where the conformally scaled metric is defined as $\tilde g=e^{-2b\alpha}g$ and the constant $b$ is defined in terms of the BD parameter $\omega $ as
\be
b=\left(\frac{2}{2\omega+3}\right)^{1/2}.
\ee
In the Einstein frame the gravitational part assumes the familiar form while coupling to the matter becomes nonminimal
while in the Jordan frame coupling to the matter is through a dynamical BD scalar field.
As a consequence, contrary to a matter energy-momentum forms relative to the Jordan frame, a matter energy-momentum is not covariantly constant in the Einstein frame. On the other hand, the construction of the expression for the BD superpotential is not modified  by a matter field
Lagrangian coupled to the BD Lagrangian.

The familiar Sparling-Thirring superpotential form can be adopted for the field equations (\ref{FeqnEinsfr}) defined  relative to the Einstein frame.
The above definition of BD superpotential (\ref{BD-superpotential-def}) facilitates the expressions in the Jordan frame.
By applying the conformal transformation used above to the generalized Sparling-Thirring form $*\mathcal{F}^a $ in the Jordan frame
\begin{eqnarray}
 *\mathcal{F}^a
&=&
-\frac{1}{2}\left(\phi \omega_{bc}+\frac{1}{2}\left(e_b i_cd\phi-e_c i_bd\phi\right) \right)\wdg *e^{abc}
\nonumber
\\
 &=&-\frac{1}{2}\phi\tilde \omega_{bc}\wdg* e^{abc},
\end{eqnarray}
where $\tilde \omega_{ab}$ is given by (\ref{tildeomega}).
Then one immediately finds
\be
\Lambda_{ab}
=
\phi \,\tilde \omega_{ab}.
\ee
Finally, by using the defining relations for the conformal transformations (\ref{ct-def}), one ends up with the remarkable result
\be
*\mathcal{F}^a
=
\phi^{1/2}\tilde* \tilde F^a,
\ee
where  $\tilde F^a$ is the Sparling-Thirring form in the Einstein frame of the form $\tilde F^a=-1/2 \, \tilde\omega_{bc}\wedge\tilde * \tilde e^{abc}$ similar to (\ref{thirring-form-def}). This equation relates the Sparling-Thirring form in the Einstein frame and the generalized BD superpotential defined in the corresponding Jordan frame.

\section{Gravitational energy for a spherically symmetric \\ vacuum metric}

In testing the validity of the formal definition of a conserved charge that makes explicit use of the field equations, an exact solution to the field equations provides  a valuable tool in explicit applications. In the discussion below, a particular spherically symmetric, static, vacuum solution to the BD theory will be taken into account. In the so-called isotropic coordinates, the metric for the spherically symmetric vacuum solution to the BD theory has the form \cite{BD-original}
\be \label{BransI}
ds^2
=
-
f^2(r)dt\otimes dt
+
h^2(r)\delta_{ij}dx^i\otimes dx^j,
\ee
with the explicit forms of the metric functions given by
\begin{eqnarray}
f(r)
&=&
e^{\alpha_0}\left(\frac{r-B}{r+B}\right)^{1/\lambda},
\\
h(r)
&=&
e^{\beta_0}\left(1+\frac{B}{r}\right)^2\left(\frac{r-B}{r+B}\right)^{(\lambda-C-1)/\lambda}. \label{h}
\end{eqnarray}
The metric functions depend only on the function $r$  defined by $r^2=\delta_{ij}x^ix^j$ in terms of the
the spatial coordinates $\{x^i\}$, for $i,j=1,2,3$.
$\alpha_0, \beta_0, B$ are integration constants and the remaining constants $C$ and $\lambda$ are related further  by
\be\label{lambdaBrel}
\lambda^2
\equiv
(C+1)^2-C(1-\frac{1}{2}\omega C).
\ee
As a function of $r$, the BD scalar is given by
\be\label{phi}
\phi=
\phi_0
\left(
\frac{r-B}{r+B}
\right)^{C/\lambda}.
\ee

The solution above, known as the Brans-I  solution in the literature, is not the only spherically symmetric, static, vacuum solution to the BD theory. There are four classes of such solutions that are usually  named as  Brans I-IV solutions \cite{Sph-sym-sol-brans-II}. However, the Brans-II solution is not an independent solution and can be obtained by a complex substitution of the parameters $C$ and $\lambda$ by making use of the Brans-I solution \cite{BhadraSarkar}.
The Brans-III  and Brans-IV solutions are  not independent of each other as well \cite{BhadraNandi}. These particular solutions require the BD parameter $\omega$ to take negative values ($\omega<-3/2$) which implies the violation of the weak energy condition. Moreover, by relating the parameters of the Brans-I solution  to the BD parameter $\omega$, it was shown in \cite{BhadraSarkar} that only the Brans-I solution can describe the gravitational field exterior to a nonsingular, spherically symmetric object obtained by matching conditions in the weak field regime. Henceforth, only the Brans-I solution given by Eqs. (\ref{BransI})-(\ref{phi}) are taken into the account in the application  below.

The explicit expressions for the Levi-Civita connection 1-forms  relative to the orthonormal coframe defined by
$
e^0
=
fdt$ and $e^k=h\, dx^k$
are given by
\begin{eqnarray}
\omega^{0j}
&=&
\frac{f'}{fr}x^j\,dt,
\label{conn-0j}\\
\omega^{jk}
&=&
-\frac{h'}{hr}
\left(
x^j\,dx^k-x^k\,dx^j
\right),\label{conn-jk}
\end{eqnarray}
where  $'$ stands for the derivative with respect to $r$.
The expansion of connection 1-forms into the associated coordinate basis coframe forms in  (\ref{conn-0j})-(\ref{conn-jk}), rather then the orthonormal coframe is convenient  in order to evaluate  the flux integral in what follows.

Moreover, with the assumption $\phi=\phi(r)$, one finds that the modified connection 1-form $\Lambda^{jk}$ has the explicit expression
\be\label{modified-conn-expression}
\Lambda^{jk}
=
-(\phi h'+\frac{1}{2}h\phi')
\frac{1}{r}
\left(
x^jdx^k-x^kdx^j
\right).
\ee

Consequently, the expression in (\ref{modified-conn-expression}) can be used in the definition (\ref{4-mom-def}) to calculate the total energy $E_{BD}$ as
\begin{eqnarray}
E_{BD}
&=&
-\lim_{r\rightarrow \infty}\frac{1}{2}\int_{S^2_r}\epsilon^{0}_{\fant{a}ijk}\Lambda^{ij}\wdg e^k\nonumber \\
&=&
\lim_{r\rightarrow \infty}\frac{1}{2}\int_{S^2_r}\epsilon_{ijk}\Lambda^{ij}\wdg e^k,
\end{eqnarray}
where $\epsilon^{0ijk}=-\epsilon_{0ijk}$ and $\epsilon_{ijk}$ stands for the  permutation symbol on the spatial submanifold $t=\mbox{constant}$. By considering (\ref{modified-conn-expression}), the total energy becomes
\be
E_{BD}
=
-\lim_{r\rightarrow \infty}\int_{S^2_r} \phi^{1/2}(\phi^{1/2}h)'
\epsilon_{ijk} \frac{x^i}{r}dx^j\wdg dx^k.
\ee

The integration measure above can be put into a form convenient for the spherical symmetry by recalling the basic geometrical formula for $S^2$ on Euclidean space $\mathcal{R}^3$. On a three dimensional Euclidian space $\mathcal{R}^{3}$, the  unit $S^2$ can simply  be defined by setting  $r=1$. In terms of the Cartesian coordinates $\{x^i\}$, one has  $r^2=\delta_{ij}x^ix^j$. By differentiating  this expression and taking the Hodge  dual one finds
\be
\star rdr
=
\frac{1}{2}\epsilon_{ijk}x^i dx^j\wdg dx^k,
\ee
where $\star$ is the  Hodge dual in $\mathcal{R}^{3}$. Restriction of the expression on the right-hand side to $S^2_r$ gives the volume element on $S^2_r$  which   is  conveniently taken to be $r^2d\Omega$ with $d\Omega$ standing for  an  infinitesimal  element of the solid angle.
Eventually, the energy expression becomes
\be\label{E-bd-form1}
E_{BD}
=-
\lim_{r\rightarrow \infty}\int_{S^2_r} 2\phi^{1/2}(\phi^{1/2}h)'r^2d\Omega.
\ee
For $\phi\sim \phi_0=$constant, the expression on the right-hand side of Eq. (\ref{E-bd-form1}) becomes proportional  to the corresponding expression in GR (see, for example, the expression given in \cite{thirring}) up to a constant.

Evaluating (\ref{E-bd-form1}) for the Brans-I solution  defined by  Eqs. (\ref{BransI})-(\ref{phi}) above,
one finds
\be \label{MBD}
E_{BD}=8 \pi M_{BD}=\frac{8\pi}{\lambda}B(2+C)\phi_0e^{\beta_0}.
\ee
The expression on the right-hand side contains five parameters, but due to the relation (\ref{lambdaBrel}), only four of them are free.

Let us compare this expression with corresponding expression of GR.
By considering the  linear expansion of the static vacuum BD solution given in (\ref{BransI}) and the scalar field (\ref{phi}), and subsequently matching
it to a static Newtonian source, the parameters $B,C,\lambda,\phi_0,\beta_0$  can be expressed  in terms of the corresponding GR parameters together with the BD parameter $\omega$ as \cite{BD-original,BhadraSarkar}
\begin{eqnarray}
 &B=\frac{\lambda M}{2},\quad C=-\frac{1}{\omega+2},\qquad \quad \beta_0=0,  \\
& \lambda=\left(\frac{2\omega+3}{2\omega+4}\right)^{1/2},
\quad \phi_0=\frac{2\omega+4}{2\omega+3}\frac{1}{G}.
\end{eqnarray}
Note that in our notation we have set $G=1.$ Putting these into (\ref{MBD}), we have found that the Einstein limit of the total energy  is given by
\be
E_E=8\pi M.
\ee
 It is possible to obtain  this result  in a more straightforward way as well. The choice $C=0,\lambda=1$ brings the Brans-I solution to the corresponding one in GR, which yields the result
\be
(E_{BD})_{\lim\omega\mapsto \infty}=16\pi  B_E
\ee
in the Einstein limit defined by  $\omega\mapsto\infty$.
In the expression above, $B_E$ is a constant and the particular choice  $B_E=M/2$ yields the correct GR limit.

\section{Concluding remarks}
Both Abbott-Deser \cite{abbott-deser} and  Deser-Tekin  \cite{deser-tekin-construction} charge definitions   make essential use of the field equations  as well as  the Killing symmetries of the background solution. For example, the Deser-Tekin charges are constructed by linearizing  the field  equations around a background solution (flat Minkowski space in our case above). Subsequently, a conserved quantity is expressed as  a flux integral in the background by making use of   the linearized equations. The construction of the generalized Sparling-Thirring superpotential is achieved  in the same spirit as the construction of a Deser-Tekin charge and the use of exterior forms and  Stokes's theorem allows one to obtain flux integrals on background spacetime in a practically useful form see, for example, the construction in  \cite{cebeci-tekin-sarioglu} by making use the Killing vector fields of the background spacetime. As for the BD example above,  the energy  in the Deser-Tekin approach is simply obtained by linearizing the generalized superpotential $*\mathcal{F}^a$.

As another important technical point, note that the construction of the BD superpotential is insensitive to a possible potential term  for the scalar field. The definition also covers the special case where the kinetic term for the BD scalar  is absent and therefore the construction  can be applied to more general $f(R)$ models since they are known to have scalar-tensor equivalent models with a potential term for the scalar field.
To facilitate a comparison with these theories, let us consider the simplest  modification of the Einstein-Hilbert Lagrangian form
\be
L=
\frac{1}{2}f(R)*1,
\ee
with $f(R)$ assumed to be an arbitrary differentiable and nonlinear function of the scalar curvature $R$.
Particular forms of the function $f$ are studied with different  motivations, for example, arising from cosmological applications.

The metric field equations for vacuum that follow from the modified Lagrangian above take the form \cite{baykal-delice}
\be\label{fr-eqn}
f'*G^a-D*(df'\wdg e^a)+\frac{1}{2}(Rf'-f)*e^a=0.
\ee
By comparing the field  equations with the BD field equation (\ref{BD-eqns1}) above one can see that the field redefinition $f'\equiv \phi$
which entails the potential term $V(\phi)$ by the Legendre transformation
\be
V(\phi)
\equiv
R(\phi)f'(R(\phi))-f(R(\phi)),
\ee
for the scalar field $\phi$. Consequently, in terms of the scalar field $\phi$, the scalar-tensor equivalent equations for (\ref{fr-eqn}) become
\be
\phi*G^a-D*(d\phi\wdg e^a)+\frac{1}{2}V(\phi)*e^a=0,
\ee
which are in fact the vacuum BD equations with $\omega=0$ and with the potential term for scalar field. The form of the field equations (\ref{fr-eqn}) is not very common in the vast literature involving various  $f(R)$ models and reader is referred to \cite{baykal-epjp} in relating (\ref{fr-eqn}) to the more familiar coordinate expression.
In either form,  the metric equations allow one to define gravitational energy in parallel to the BD case presented above immediately. By using the form of the field equation  given in (\ref{fr-eqn}), one finds the total energy by the flux integral
\begin{eqnarray}
E_{f(R)}
\equiv&&
-\frac{1}{2}\int_{\partial U}\left[f'\,\omega_{bc}
\, +\frac{1}{2}f''\left(e_b i_cdR-e_c i_bdR \right)\right]\wdg *e^{0bc}.
\end{eqnarray}
For the simple case $f(R)=R$ the above formula gives back the Sparling-Thirring form up to a constant multiple. For the next simple case
with $f(R)=R+\alpha R^2$
the leading term is the second term on the right-hand side and
the above formula reduces up to a constant multiple to   the energy definition  of Deser-Tekin  for a particular subcase of general quadratic curvature gravity in four dimensions \cite{deser-tekin-construction}. It follows by definition that the energy has the expression for the $R^2$ term is of the form
\be
E_{R^2}
\sim
\int_{\partial U}*(d R\wdg e^0)
\ee
for asymptotically flat solutions \cite{baykal-prd}.
Further scrutiny of the generalized Sparling-Thirring  forms for modified gravity models in relation to the definitions of the
Deser-Tekin  charges will be taken up elsewhere.

\end{document}